\documentclass{ws-ijmpa}
\newcommand{\unit}{\leavevmode\hbox{\small1\kern-3.6pt\normalsize1}}

\def\lsim{\raise0.3ex\hbox{$\;<$\kern-0.75em\raise-1.1ex\hbox{$\sim\;$}}}
\def\gsim{\raise0.3ex\hbox{$\;>$\kern-0.75em\raise-1.1ex\hbox{$\sim\;$}}}

\def    \bea           	{\begin{eqnarray}}
\def    \eea           	{\end{eqnarray}}

\begin{document}

\markboth{Daniel E. Lopez-Fogliani}{The seesaw mechanism in the $\mu \nu$SSM}

%
%

\title{The seesaw mechanism in the $\mu \nu$SSM}

\author{Daniel E. L\'opez-Fogliani} 

\address{Department of Physics and Astronomy, University of Sheffield,\\ Sheffield S3 7HR, England}

\maketitle

\begin{abstract}

The $\mu \nu$SSM  proposes to use  right-handed neutrino supermultiplets in order to  generate  the $\mu$ term and neutrino masses simultaneously. We discuss  neutrino physics and the associated electroweak seesaw mechanism in this model. We  show how to obtain, from  the neutralino-neutrino mass matrix of the $\mu \nu$SSM, the effective neutrino mass matrix. In particular we discuss certain limits of this matrix that clarify the neutrino-sector behavior of the model.  We also show that current data on neutrino masses and mixing angles can easily be reproduced.  These constraints can be fulfilled even with a diagonal neutrino Yukawa matrix, since this seesaw does not involve only the right-handed neutrinos but also the MSSM neutralinos. To obtain the correct neutrino angles turns out to be easy  due to the following characteristics of this seesaw: R-parity is broken and the relevant scale is the electroweak one.


\end{abstract}



\section{Introduction}
The ``$\mu$ from $\nu$'' Supersymmetric Standard Model ($\mu\nu$SSM) was proposed
in Ref. \refcite{MuNuSSM} as an alternative to the Minimal Supersymmetric Standard Model (MSSM). In particular, it provides a solution to the $\mu$-problem \cite{muproblem} of the MSSM and explains the origin of neutrino masses by simply using right-handed neutrino superfields. 

Since it was proposed \cite{MuNuSSM}, the  $\mu\nu$SSM has been studied in Ref. \refcite{MuNuSSM2}
-\refcite{Ghosh2010}
(for a summary see Ref. \refcite{carlosresumen}). The superpotential of the  $\mu\nu$SSM is given by\cite{MuNuSSM}:

\bea\label{superpotential}
W  &= &
\ \epsilon_{ab} \left(
Y_u^{ij} \, \hat H_u^b\, \hat Q^a_i \, \hat u_j^c +
Y_d^{ij} \, \hat H_d^a\, \hat Q^b_i \, \hat d_j^c +
Y_e^{ij} \, \hat H_d^a\, \hat L^b_i \, \hat e_j^c +
Y_\nu^{ij} \, \hat H_u^b\, \hat L^a_i \, \hat \nu^c_j 
\right)\nonumber \\ 
&-&\epsilon_{ab} \lambda^{i} \, \hat \nu^c_i\,\hat H_d^a \hat H_u^b
+
\frac{1}{3}
\kappa^{ijk} 
\hat \nu^c_i\hat \nu^c_j\hat \nu^c_k\,.
\eea

A $Z^3$ symmetry could be evoked
to forbid parameters with dimensions in the superpotential.
This symmetry may have a clear origin if the Lagrangian  describes massless modes of a fundamental object (with very massive modes, as in string theory).
Non-renormalizable contributions  could lift this $Z^3$  symmetry avoiding a possible domain wall problem without introducing hierarchy problems, in a similar way as in the NMSSM case \cite{abel}.

Working in the framework of gravity mediated supersymmetry breaking,
the following soft terms appear in the Lagrangian~\cite{MuNuSSM},

\bea
-\mathcal{L}_{\text{soft}} & = &
 (m_{\tilde{Q}}^2)^{ij} \, \tilde{Q^a_i}^* \, \tilde{Q^a_j}
+(m_{\tilde u^c}^{2})^{ij} 
\, \tilde{u^c_i}^* \, \tilde u^c_j
+(m_{\tilde d^c}^2)^{ij} \, \tilde{d^c_i}^* \, \tilde d^c_j 
+(m_{\tilde{L}}^2)^{ij} \, \tilde{L^a_i}^* \, \tilde{L^a_j}
+(m_{\tilde e^c}^2)^{ij} \, \tilde{e^c_i}^* \, \tilde e^c_j
\nonumber \\
&+ &
m_{H_d}^2 \,{H^a_d}^*\,H^a_d + m_{H_u}^2 \,{H^a_u}^* H^a_u +
(m_{\tilde\nu^c}^2)^{ij} \,\tilde{{\nu}^c_i}^* \tilde\nu^c_j 
\nonumber \\
&+&
\epsilon_{ab} \left[
(A_uY_u)^{ij} \, H_u^b\, \tilde Q^a_i \, \tilde u_j^c +
(A_dY_d)^{ij} \, H_d^a\, \tilde Q^b_i \, \tilde d_j^c +
(A_eY_e)^{ij} \, H_d^a\, \tilde L^b_i \, \tilde e_j^c 
\right.
\nonumber \\
&+&
\left.
(A_{\nu}Y_{\nu})^{ij} \, H_u^b\, \tilde L^a_i \, \tilde \nu^c_j 
- (A_{\lambda}\lambda)^{i} \, \tilde \nu^c_i\, H_d^a  H_u^b
+ \text{H.c.}
\right] 
+ \left[
\frac{1}{3}
(A_{\kappa}\kappa)^{ijk} 
\tilde \nu^c_i \tilde \nu^c_j \tilde \nu^c_k\
+ \text{H.c.} \right]
\nonumber \\
&-&  \frac{1}{2}\, \left(M_3\, \tilde\lambda_3\, \tilde\lambda_3+M_2\,
  \tilde\lambda_2\, \tilde
\lambda_2
+M_1\, \tilde\lambda_1 \, \tilde\lambda_1 + \text{H.c.} \right) \,.
\label{2:Vsoft}
\eea
%
Let us recall that the only scale in the model is the SUSY breaking scale present in the above soft terms~\cite{MuNuSSM}.

%

Once the electroweak symmetry is spontaneously broken, the neutral scalars develop in general the following VEVs, $\, \langle H_d^0 \rangle = v_d \, , \quad \langle H_u^0 \rangle = v_u \, , \quad \langle \tilde \nu_i \rangle = \nu_i \, , \quad  \langle \tilde \nu_i^c \rangle = \nu_i^c \,$.
%
An effective $\mu$ term, $\mu \equiv \lambda_i \nu_i^c$, is generated from  $\lambda^{i} \, \hat \nu^c_i\,\hat H_d \hat H_u$ in the superpotential. In a similar way  terms of the type $\hat \nu^c\hat \nu^c\hat \nu^c\,$ are generating an effective Mayorana mass  $\sim \kappa \: \nu_i^c$ of  electroweak order. Note that  the presence at the same time of the Yukawa term for the neutrino and the last two terms in (\ref{superpotential}) are breaking R-parity explicitly.


Since in this model  R-parity is explicitly broken, the neutralino  is not a possible candidate for dark matter (and neither is the sneutrino). In this respect the  gravitino as a dark matter candidate was studied in Ref. \refcite{gravitino}, where a possible detection was discussed.

The breaking of R-parity can easily be understood if we realize that in the limit where Yukawas for neutrinos are vanishing, the $\hat \nu^c$ are just ordinary singlet superfields, without any connection with neutrinos, and this model would coincide (although with three instead of one singlet) with the Next-to-Minimal Supersymmetric Standard Model (NMSSM) where R-parity is conserved.

Once we switch on the neutrino Yukawa couplings, the fields $\hat \nu^c$ become right-handed neutrino superfields, and, as a consequence, R-parity is broken. Indeed this breaking is small because
we have an electroweak-scale seesaw, implying neutrino Yukawa couplings no larger than $10^{-6}$ (like the electron Yukawa) \cite{MuNuSSM}.


Note that in principle the number of right-handed neutrinos 
is a free parameter. To fix the number
we have  followed in this work the philosophy  that  three families for all leptons is natural.
With respect to this, the  characteristics of  the $\mu \nu$SSM with different number of neutrinos have been discusses  in Ref. \refcite{Hirsch0}, where the  case of only one right-handed neutrino is specially discussed. Regarding the case of two right-handed neutrinos,  which is the minimal number to generate the experimental pattern at tree level, 
it was pointed out in Ref. \refcite{neutrinos} that two is also the minimal number to have the possibility of Spontaneous CP Violation (SCPV).


%
%

In the following we describe the  seesaw mechanism  in the $\mu \nu$SSM.



\section{The seesaw mechanism in the $\mu \nu$SSM \label{section:neutrino}}

In the $\mu \nu$SSM the MSSM neutralinos mix with the left- and right-handed neutrinos as a consequence of R-parity violation. Therefore  the right-handed neutrinos behave as singlino components of the neutralinos. In the basis 
${\chi^{0}}^T=(\tilde{B^{0}},\tilde{W^{0}},\tilde{H_{d}},\tilde{H_{u}}, 
        \nu_{R_i},\nu_{L_i})$ 
the neutralino-neutrino mass matrix was given in Refs. \refcite{MuNuSSM} and \refcite{MuNuSSM2} for real VEVs, and in Ref. \refcite{neutrinos} for complex ones. For simplicity we work in the real case, and then  we have,
\begin{align}
{\cal M}_n=\left(\begin{array}{cc}
M & m\\
m^{T} & 0_{3\times3}
\label{Mneutralinos10x10}
\end{array}\right),
\end{align} 
where the  mass matrix, $M$, is,
{\small \begin{align}
\left(
\begin{array}{ccccccc}
M_{1} & 0 & -A \langle H_d^0 \rangle & A \langle H_u^0 \rangle & 0 & 0 & 0\\
0 & M_{2} &B \langle H_d^0 \rangle & -B \langle H_u^0 \rangle & 0 & 0 & 0\\
-A \langle H_d^0 \rangle & B \langle H_d^0 \rangle & 0 & 
-\lambda_{i}\langle \tilde \nu_i^c \rangle  & 
-\lambda_{1}\langle H_u^0 \rangle & -\lambda_{2}\langle H_u^0 \rangle & 
-\lambda_{3}\langle H_u^0 \rangle\\
A \langle H_u^0 \rangle & -B \langle H_u^0 \rangle & 
-\lambda_{i}\langle \tilde \nu_i^c \rangle  & 0 & 
-\lambda_{1}\langle H_d^0 \rangle + Y_{\nu_{i1}}\langle \tilde \nu_i \rangle  & 
-\lambda_{2}\langle H_d^0 \rangle + Y_{\nu_{i2}}\langle \tilde \nu_i \rangle  & 
-\lambda_{3}\langle H_d^0 \rangle + Y_{\nu_{i3}}\langle \tilde \nu_i \rangle \\
0 & 0 & -\lambda_{1}\langle H_u^0 \rangle & 
-\lambda_{1}\langle H_d^0 \rangle + Y_{\nu_{i1}}\langle \tilde \nu_i \rangle  & 
2\kappa_{11j}\langle \tilde \nu_j^c \rangle  & 
2\kappa_{12j}\langle \tilde \nu_j^c \rangle  & 
2\kappa_{13j}\langle \tilde \nu_j^c \rangle \\
0 & 0 & -\lambda_{2}\langle H_u^0 \rangle & 
-\lambda_{2}\langle H_d^0 \rangle + Y_{\nu_{i2}}\langle \tilde \nu_i \rangle  & 
2\kappa_{21j}\langle \tilde \nu_j^c \rangle  & 
2\kappa_{22j}\langle \tilde \nu_j^c \rangle  & 
2\kappa_{23j}\langle \tilde \nu_j^c \rangle \\
0 & 0 & -\lambda_{3}\langle H_u^0 \rangle & 
-\lambda_{3}\langle H_d^0 \rangle + Y_{\nu_{i3}}\langle \tilde \nu_i \rangle  & 
2\kappa_{31j}\langle \tilde \nu_j^c \rangle  & 
2\kappa_{32j}\langle \tilde \nu_j^c \rangle  & 
2\kappa_{33j}\langle \tilde \nu_j^c \rangle \end{array}
\right), \label{Mneutralinos7x7}
\end{align}
}
with $A = \frac{G}{\sqrt{2}} \sin\theta_W$, 
$B = \frac{G}{\sqrt{2}} \cos\theta_W$, and
\begin{align}
m^{T}=\left(\begin{array}{ccccccc}
-\frac{g_{1}}{\sqrt{2}}\langle \tilde \nu_1 \rangle  & 
\frac{g_{2}}{\sqrt{2}}\langle \tilde \nu_1 \rangle  & 
0 & 
Y_{\nu_{1i}}\langle \tilde \nu_i^c \rangle  & 
Y_{\nu_{11}}\langle H_u^0 \rangle & 
Y_{\nu_{12}}\langle H_u^0 \rangle & Y_{\nu_{13}}\langle H_u^0 \rangle\\
-\frac{g_{1}}{\sqrt{2}}\langle \tilde \nu_2 \rangle  & 
\frac{g_{2}}{\sqrt{2}}\langle \tilde \nu_2 \rangle  & 
0 & 
Y_{\nu_{2i}}\langle \tilde \nu_i^c \rangle  & 
Y_{\nu_{21}}\langle H_u^0 \rangle & 
Y_{\nu_{22}}\langle H_u^0 \rangle & 
Y_{\nu_{23}}\langle H_u^0 \rangle\\
-\frac{g_{1}}{\sqrt{2}}\langle \tilde \nu_3 \rangle  & 
\frac{g_{2}}{\sqrt{2}}\langle \tilde \nu_3 \rangle  & 
0 & 
Y_{\nu_{3i}}\langle \tilde \nu_i^c \rangle  & 
Y_{\nu_{31}}\langle H_u^0 \rangle & 
Y_{\nu_{32}}\langle H_u^0 \rangle & 
Y_{\nu_{33}}\langle H_u^0 \rangle\end{array}\right). \label{m7x3} \end{align}
The above matrix (\ref{Mneutralinos10x10}) is of the seesaw type 
where the entries of $M$ are of the order of the electroweak scale while the ones in $m$ are of the order of the Dirac masses for the neutrinos \cite{MuNuSSM,MuNuSSM2}. Therefore  in a first approximation the effective neutrino mixing mass matrix can be written as,
\begin{equation}
m_{eff} = -m^T \cdot M^{-1} \cdot m, \label{eff}
\end{equation}
and one can diagonalize it by a unitary transformation (ortonormal in the real case)
\begin{equation}
U_{MNS}^T m_{eff} U_{MNS} = diag(m_{\nu_1}, m_{\nu_2}, m_{\nu_3}).
\label{nudagnudiag}
\end{equation}


Let us write an approximate analytical expression for the effective neutrino mass matrix of the $\mu \nu$SSM
neglecting all the terms containing $Y_{\nu}^2 \nu^2$, $Y_{\nu}^3 \nu$ and $Y_{\nu} \nu^3$,  due to the smallness of $Y_{\nu}$ and $\nu$ \cite{MuNuSSM}, and taking couplings $\lambda_i \equiv \lambda$, the tensor $\kappa$ with terms $\kappa_{iii} \equiv \kappa_i \equiv \kappa$ and vanishing otherwise, diagonal Yukawa couplings $Y_{\nu_{ii}} \equiv Y_{\nu_i}$, and VEVs $\nu_i^c\equiv\nu^c$. From Ref. \refcite{neutrinos} (also this coincides with the results of Ref. \refcite{Ghosh:2008yh}) we have,

\begin{eqnarray}
\label{Formula analitica real reescrita}
& (m_{eff})_{ij} \simeq \frac{v_u^2}{6\kappa \nu^c}Y_{\nu_i}Y_{\nu_j}
                     \left(1-3 \, \delta_{ij}\right)  
&  -\frac{1}{2M_{eff}}\left[\nu_i \nu_j+\frac{v_d\left(Y_{\nu_i}\nu_j
   +Y_{\nu_j}\nu_i\right)}{3\lambda}
   +\frac{Y_{\nu_i}Y_{\nu_j}v_d^2}{9\lambda^2 }\right]\ ,
   \nonumber\\
  \end{eqnarray}
with
\begin{eqnarray}
 M_{eff}\equiv M \left[1-\frac{v^2}{2M \left(\kappa \nu^{c^2}+\lambda v_u v_d\right)
        \ 3 \lambda \nu^c}\left(2 \kappa \nu^{c^2} \frac{v_u v_d}{v^2}
        +\frac{\lambda v^2}{2}\right) \right]\ .
\end{eqnarray}
Here
$v^2=v_u^2+v_d^2+\sum_i\nu_i^2 \approx v_u^2+v_d^2 \approx (174 \text{GeV})^2 $
has been used, since $\nu_i <<v_u,v_d$ \cite{MuNuSSM},
and also we have defined $\frac{1}{M} = \frac{g_1^2}{M_1}+\frac{g_2^2}{M_2}$.

To clarify how the seesaw works in this model let us make three limits, making special emphasis in the last one.

The first limit \cite{Ghosh:2008yh,neutrinos} is  the one where gauginos are very heavy and decouple
(i.e. $M \rightarrow \infty$), and then  (\ref{Formula analitica real reescrita}) 
reduces to
\begin{eqnarray}
(m_{eff})_{ij} \simeq 
         \frac{v_u^2}{6 \, \kappa \nu^c}Y_{\nu_i}Y_{\nu_j}
\left(1-3 \, \delta_{ij}\right).
\label{Not Ordinary see-saw}
\end{eqnarray}
The effective mixing of the right-handed neutrinos and Higgsinos in this limit produces off-diagonal entries in the mass matrix. Besides, when right-handed neutrinos 
are also decoupled (i.e. $\nu^c \to \infty $), the neutrino masses 
are zero as corresponds to the case of a seesaw with only Higgsinos.


The second limit \cite{Ghosh:2008yh,neutrinos} which is worth discussing is 
$\nu^c \rightarrow \infty$. Then,  
(\ref{Formula analitica real reescrita}) reduces to the form 
\bea
(m_{eff})_{ij} \simeq-\frac{1}{2M}\left[\nu_i \nu_j
      +\frac{v_d(Y_{\nu_i}\nu_j+Y_{\nu_j}\nu_i)}{3\lambda}
      +\frac{Y_{\nu_i}Y_{\nu_j}v_d^2}{9\lambda^2 }\right].
\eea
We can also see that for $v_d \rightarrow 0$ (i.e. 
$\tan \beta=\frac{v_u}{v_d}\to \infty$) 
one obtains
\begin{eqnarray}
(m_{eff})_{ij} \simeq -\frac{\nu_i \nu_j}{2M}.
\label{Gaugino see-saw}
\end{eqnarray}
Note that this result can actually be obtained if $\nu_i >> \frac{Y_{\nu_i} v_d}{3 \lambda}$, as was noticed in Ref. \refcite{neutrinos}.
It means that decoupling right-handed neutrinos/singlinos and 
Higgsinos, the seesaw mechanism is generated through the mixing of 
left-handed neutrinos with gauginos. This is a characteristic feature 
of the seesaw in the well-known bilinear R-parity violation model (BRpV) \cite{Cita44paperHuitu}.
Notice that the gauginos (Bino and Wino) are in the adjoint representation, so they have a vector character, and (\ref{Gaugino see-saw}) could be associated to a type III seesaw \footnote{The association of (\ref{Gaugino see-saw}) to a type III seesaw was pointed out in Ref.  \refcite{Ghosh2010}.} as (\ref{Not Ordinary see-saw}) is associated to a  Type I.
 Nevertheless  we recall that there is an intrinsically supersymmetric character of this seesaw, since gauginos are part of gauge supermultiplets. 

The last limit \cite{neutrinos} we want to discuss gives a clear idea of the seesaw mechanism in this model.
The seesaw in the $\mu\nu$SSM comes, in general, from the interplay of the above two 
limits. Namely, the limit where we suppress only certain Higgsino and 
gaugino mixing. Hence, taking $v_d \rightarrow 0$ in  (\ref{Formula analitica real reescrita}),
we obtain
\begin{eqnarray}
(m_{eff})_{ij}\simeq   \frac{v_u^2}{6 \kappa \nu^c}Y_{\nu_i}Y_{\nu_j}
         (1-3 \delta_{ij})-\frac{1}{2 \, M_{\text{eff}}}\nu_i \nu_j\ ,
\label{Limit quasi no mixing Higgsinos gauginos}
\end{eqnarray}
As above, we remark that actually this result can be obtained 
if $\nu_i >>\frac{Y_{\nu_i} v_d}{3 \lambda}$.
The effective mass
$M_{\text{eff}} = M \left(1-\frac{v^4}{12 \kappa M  \nu^{c^3}  }\right)$
represents the mixing between gauginos and Higgsinos-$\nu^c$ that is not completely suppressed in this limit.
Expression (\ref{Limit quasi no mixing Higgsinos gauginos}) is more general than the other two limits studied above.
Notice that for typical values of the parameters involved in the seesaw,$M_{\text{eff}}\approx M$, and therefore we get a simple formula that can be used to understand the seesaw mechanism in this model in an qualitative way, that is
\begin{eqnarray}
(m_{eff})_{ij}\simeq \frac{v_u^2}
{6 \kappa \nu^c}Y_{\nu_i}Y_{\nu_j}
                   (1-3 \delta_{ij})-\frac{1}{2M} \nu_i \nu_j.
\label{Limit no mixing Higgsinos gauginos}
\end{eqnarray}
The simplicity of  (\ref{Limit no mixing Higgsinos gauginos}), in contrast with the full formula given by  (\ref{Formula analitica real reescrita}), comes from the fact that the mixing between gauginos and Higgsinos-$\nu^c$ is neglected. In this model the effective neutrino mass matrix is not diagonal even for diagonal Yukawas.  Following Ref. \refcite{neutrinos}, in the next section we  show how to obtain the normal and inverted hierarchy, reproducing experimental masses and mixing angles, in an easy way, using diagonal Yukawas  \footnote{The first solutions for masses and mixing angles in the experimental allowed range using diagonal Yukawas were found in Ref. \refcite{Ghosh:2008yh}.}.

\section{Normal and Inverted hierarchies obtained with diagonal Yukawas}

\begin{figure}[ht]
\centerline{\psfig{file=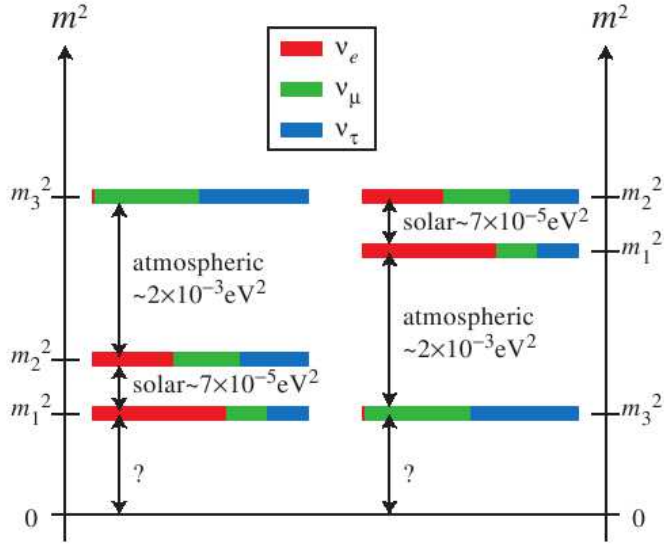,width=10cm}}
\vspace*{8pt}
\caption{The two possible hierarchies of neutrino masses. The pattern on the left side corresponds to the  normal hierarchy and is characterized by one heavy state with a very little electron neutrino component, and two light states with a mass difference which is the solar mass difference. The pattern on the right side corresponds to the inverted hierarchy and is characterized by two heavy states with a mass difference that is the solar mass difference, and a light state which has very little electron neutrino component.}\label{ImagenJerarquias}
\end{figure}

To continue the discussion of the seesaw in the
$\mu\nu$SSM, let us  remind that two mass differences and mixing angles have been measured 
experimentally in the neutrino sector. The allowed $3 \sigma$ ranges for these parameters are shown in Table \ref{Synopsis} (as discussed in Ref. \refcite{ConstraintsFogli}).
We also show the compositions of the mass eigenstates in Fig.~\ref{ImagenJerarquias} for the normal and inverted hierarchy cases (as shown in Ref. \refcite{Dibujo Jerarquias}). 

\begin{table}[hb]
\tbl{Allowed $3 \sigma$ ranges for the neutrino masses and mixings.}
{\begin{tabular}{|c|c|c|c|c|}
\hline
$\Delta m_{sol}^2/10^{-5}\mathrm{\ eV}^2$ & $\sin^2\theta_{12}$ & 
            $\sin^2\theta_{13}$ & $\sin^2\theta_{23}$ &
            $\Delta m_{atm}^2/10^{-3}\mathrm{\ eV}^2$ \\[5pt]
\hline
7.14-8.19 & 0.263-0.375  & $<0.046$        & 
0.331-0.644  & 2.06-2.81 \\
\hline
\end{tabular} \label{Synopsis}}
\end{table}

Due to the fact that the mass eigenstates have, in a good  approximation, the same composition of $\nu_{\mu}$ and $\nu_{\tau}$ we start considering  diagonal and degenerate Yukawas in $\mu$ and $\tau$,  $Y_{\nu_2}=Y_{\nu_3}$,  also with $\nu_2=\nu_3$. Therefore  (\ref{Limit no mixing Higgsinos gauginos}) takes the form
\begin{align}
m_{eff}=\left(
\begin{array}{ccc}
d & c & c \\
c & A & B \\
c & B & A
\end{array}\right),
\label{meff toy model}
\end{align}

where
\begin{eqnarray}
d=-\frac{v_u^2}{3 \kappa \nu^c}Y_{\nu_1}^2-\frac{1}{2M}\nu_1^2, \; \; 
c=\frac{v_u^2}{6\kappa \nu^c}Y_{\nu_1}Y_{\nu_2}
                            -\frac{1}{2M}\nu_1 \nu_2, \nonumber \\
A=-\frac{v_u^2}{3 \kappa \nu^c}Y_{\nu_2}^2-\frac{1}{2M}\nu_2^2, \; \; 
B=\frac{v_u^2}{6\kappa \nu^c}Y_{\nu_2}^2-\frac{1}{2M}\nu_2^2.
\label{Parameters toy model}
\end{eqnarray}
The eigenvalues of this matrix are the following:
\begin{eqnarray}
\frac{1}{2}\left(A+B-\sqrt{8c^2+(A+B-d)^2}+d \right),
  \frac{1}{2}\left(A+B+\sqrt{8c^2+(A+B-d)^2}+d \right), 
  A-B\ ,                                    
\nonumber \\
\label{Eigenvalues toy model}
\end{eqnarray}
and the corresponding eigenvectors (for simplicity are not normalised) 
are
\begin{eqnarray}
\left(-\frac{A+B+\sqrt{8c^2+(A+B-d)^2}-d}{2},c,c \right),  \; 
& \left(\frac{-A-B+\sqrt{8c^2+(A+B-d)^2}+d}{2c},1,1 \right),  \; 
& \left(0,-1,1 \right).\;  
\label{Eigenvectors toy model}
\end{eqnarray}
We have ordered the eigenvalues in such a way that it is clear how to 
obtain the normal hierarchy for the $\nu_{\mu}$-$\nu_{\tau}$ degenerate case. Then we see that $\sin^2\theta_{13}=0$ and $\sin^2\theta_{23}=\frac{1}{2}$, as in the bimaximal mixing regime. Also we have enough freedom to fix the parameters in such a way that the experimental values for the mass differences and the remaining angle $\theta_{12}$ can be reproduced.
It is important to mention that the above two values of the angles 
are a consequence of considering the example with $\nu_{\mu}$-$\nu_{\tau}$ degeneration, and therefore valid even if we use the general 
formula (\ref{Formula analitica real reescrita}) instead of the simplified expression 
(\ref{Limit no mixing Higgsinos gauginos}). 
Notice that  (\ref{meff toy model}), (\ref{Eigenvalues toy model}) and (\ref{Eigenvectors toy model}) would be the same but with the corresponding
values of $A, B, c$ and $d$.

Let us finally remark that we can get the  tri-bimaximal mixing regime, i.e., 
$\sin^2\theta_{13}= 0$, $\sin^2\theta_{23}=1/2$ and $\sin^2\theta_{12}=1/3$, fixing in  (\ref{meff toy model}) $c=A+B-d$. In this way we obtain the eigenvalues
\begin{eqnarray}
-(A+B)+2d \ , \ 2(A+B)-d, \ A-B.
\label{Eigenvalues toy model c to 0ppp}
\end{eqnarray}

It is important to note that, in this case, we need $|A-B|>|2(A+B)-d|$ for the normal hierarchy 
case, otherwise the $\theta_{12}$ angle is zero instead of $\theta_{13}$.

In the inverted hierarchy scenario $|A-B|>|2(A+B)-d|$ leads the angle $\theta_{12}$ to zero  which is not phenomenologically viable. Then we impose $|A-B|<|2(A+B)-d|$.

In this way we obtain the eigenvalues,
\begin{eqnarray}
-(A+B)+2d=m_{\nu_1} \ , \ 2(A+B)-d=m_{\nu_2}, \ A-B=m_{\nu_3},
\label{Eigenvalues}
\end{eqnarray}
Where for the normal case $ |A-B|=m_{\nu_3} $ is the heaviest one and for the inverted case is the lightest one, as in Fig \ref{ImagenJerarquias}.

Breaking the degeneracy between the $Y_{\nu}$ and $\nu$ for the muon and tau neutrinos, it is possible to find more general solutions in the normal and inverted hierarchy cases. Here we have shown an example  to reproduce the experiment, but more possibilities can be found in Refs. \refcite{Ghosh:2008yh,neutrinos} and \refcite{Ghosh2010}.


It is remarkable that one could assume  diagonal Yukawas for the leptons, reproducing all the neutrinos masses and mixing angle. In other words, the leptonic sector of this model has the ability to give large mixing angles using a simple  diagonal structure for the Yukawas.  In a sense, the model gives an  explanation of why the mixing angles of  the quark and lepton sector are so different.


The characteristics of the  seesaw in this model can easily be understood from the limit  (\ref{Limit no mixing Higgsinos gauginos}): R-parity is broken (neutralinos
are part of the seesaw) and  the only scale is the SUSY one \cite{neutrinos}.


\noindent {\bf Acknowledgements}

I gratefully acknowledge C. Mu\~noz 
for useful conversations. I also wishes to thank the organizers of the `BUE, CTP International Conference on Neutrino Physics in the LHC Era' at Luxor for their hospitality.
This work was supported by the  STFC.



\begin{thebibliography}{99}

\bibitem{MuNuSSM} 
D.~E.~L\'opez-Fogliani and C.~Mu\~noz,
{\it Phys. Rev. Lett.} {\bf 97} (2006) 041801 [arXiv:hep-ph/0508297].
%

\bibitem{muproblem}
J.~E.~Kim and H.~P.~Nilles,
{\it Phys. Lett.} {\bf B138} (1984) 150.


\bibitem{MuNuSSM2}
N.~Escudero, D.~E.~L\'opez-Fogliani, C.~Mu\~noz and R.~Ruiz~de~Austri,
{\it JHEP} {\bf 12} (2008) 099
[arXiv:0810.1507 [hep-ph]].


\bibitem{neutrinos}
J. Fidalgo, D.~E.~L\'opez-Fogliani, C.~Mu\~noz and R.~Ruiz~de~Austri,
{\it JHEP} {\bf 08} (2009) 105 [arXiv:0904.3112[hep-ph]].


\bibitem{gravitino} K. Y. Choi, D. E. L\'opez-Fogliani, C. Munoz, R. Ruiz de Austri,  {\it JCAP} {\bf 1003} (2010) 028 [arXiv:0906.3681[hep-ph]]. 


\bibitem{Ghosh:2008yh}
  P.~Ghosh and S.~Roy,
  {\it JHEP} {\bf 0904} (2009) 069
  [arXiv:0812.0084 [hep-ph]].

\bibitem{Hirsch0}
  A.~Bartl, M.~Hirsch, A.~Vicente, S.~Liebler and W.~Porod,
  {\it JHEP} {\bf 0905} (2009) 120
  [arXiv:0903.3596 [hep-ph]].

\bibitem{Ghosh2010} P. Ghosh, P. Dey, B. Mukhopadhyaya, S. Roy,  ``Radiative contribution to neutrino masses and mixing in $\mu\nu$SSM,'' arXiv:1002.2705 

\bibitem{carlosresumen} C. Mu\~noz, {\it ``Phenomenology of a New Supersymmetric Standard Model: The $\mu\nu$SSM''}, arXiv:0909.5140 [hep-ph].

\bibitem{abel} 
S.~A.~Abel,
{\it Nucl.\ Phys.} {\bf B480} (1996) 55 [hep-ph/9609323];
C.~Panagiotakopoulos and K.~Tamvakis,
{\it Phys.\ Lett.} {\bf B446} (1999) 224 [hep-ph/9809475].
%


\bibitem{Cita44paperHuitu} 
M. Hirsch, M.A. Diaz, W. Porod, J.C. Romao and J.W.F. Valle, 
{\it Phys. Rev.} {\bf D62} (2000) 113008 [arXiv:hep-ph/0004115], Erratum-ibid. {\bf D65} (2000) 119901.

\bibitem{ConstraintsFogli} 
G.L. Fogli, E. Lisi, A. Marrone, A. Melchiorri, 
A. Palazzo, A. M. Rotunno, P. Serra, J. Silk and A. Slosar, 
{\it Phys. Rev.} {\bf D78} (2008) 033010 [arXiv:0805.2517 [hep-ph]].


\bibitem{Dibujo Jerarquias}
S.~F.~King,
 ``Neutrino Physics,''
arXiv:0712.1750 [physics.pop-ph].



\end{thebibliography}
\end{document}